\documentclass[9pt,twocolumn,twoside]{article}
\usepackage{amsmath,amsfonts,amssymb}
\usepackage{graphicx,xcolor}

\begin{document}

\title{Confidence intervals for the encircled energy fraction and the
  half energy width}
\author{Giuseppe Vacanti\\cosine Science \& Computing BV, Oosteinde 36, 2361 HE Warmond, The Netherlands}

\maketitle

\begin{abstract}
  The Encircled Energy Fraction and its quantiles, notably the Half
  Energy Width, are routinely used to characterize the quality of
  X-ray optical systems. They are however always quoted without a
  statistical error. We show how non-parametric statistical methods
  can be used to redress this situation, and we discuss how the
  knowledge of the statistical error can be used to speed up the
  characterization efforts for future X-ray observatories.
\end{abstract}

\section{Introduction}
\label{sec:introduction}

One of the parameters used to characterize the performance of optics
is the diameter of the region of the focal plane containing a certain
fraction of the total power transmitted by the optics. In X-ray optics
the diameter of the disk containing 50\% of the photons collected in
the focal plane is commonly used, and it is called the Half Energy
Width (HEW); at other wavelengths this is known as the Half Power
Disk.

The HEW is a key performance indicator of the angular resolution of
the optics, and very often it is the only performance indicator used
to describe the quality of the optics. Comparisons between different
X-ray mirror technologies are also often done solely in terms of
achieved HEW.

To our knowledge, nobody ever quotes the statistical error associated
with the HEW, nor any confidence band is given when the Encircled
Energy Fraction (EEF) is plotted. Still, this is a natural thing to do
when measurement results are reported. Indeed it is almost always the
case that the functional form of the point spread function of an
optics is not known, and therefore there is no ready-to-use formula
that can be applied to bound the experimental values measured. This
observation however suggests that a solution to the problem should be
looked for in non-parametric statistical methods.

In this article we show that by making use of textbook non-parametric
statistics solutions, the statistical error and the confidence interval
for a measure HEW value, and the confidence band of the EEF can be
easily determined.

The remainder of this article is organized as follows. In
\S~\ref{sec:conf-band-encircl} we describe the experimental
configuration considered, and recast the notion of EEF in terms of the
empirical Cumulative Distribution Function (CDF); once this is done
the confidence band of the EEF can be easily calculated. In
\S~\ref{sec:standard-error-half} we continue mapping terms known
in optics to terms used in probability and statistical theory; again
this will lead in a straightforward manner  to calculate the standard error of the
HEW. The logical extension to the calculation of a confidence interval
for the HEW is illustrated in \S~\ref{sec:conf-interv-half}. The
solutions discussed are illustrated by means of numerical experiments
in \S~\ref{sec:numer-exper}, and further discussed in
\S~\ref{sec:discussion}. Conclusions are drawn in
\S~\ref{sec:conclusions}.

\section{Confidence band for the encircled energy fraction}
\label{sec:conf-band-encircl}
Consider an experimental set up designed to characterize the optical
performance of an imaging system. We limit our discussion to a photon
counting experiment. Without loss of generality we can
limit ourselves to the case where the source of photons is on-axis and
monochromatic. Photons are reflected by the optical system and
collected at the focal plane with an ideal position-sensitive
detector.

For a certain measurement let $n$ be the number of photons
detected. Let $\{ R_i \}_{i=1 \ldots n}$ be the distances of the
detected photons from the center of the focal plane, the place where a
perfect imaging system would image a point source at infinity. The
empirical Cumulative Distribution Function (CDF) is defined as
\begin{equation}
\label{eq:7}
  \widehat{F}_n(r) = \frac{1}{n} \sum_1^n I(R_i\le r)\,,
\end{equation}
where
\begin{equation}
\label{eq:8}
  I(R_i \le r) = \left\{
      \begin{array}{ll}
        1 & \quad \text{if $R_i \le r$}\\
        0 & \quad \text{otherwise}\,.
      \end{array} \right.
\end{equation}
It is obvious that the CDF is equivalent to the EEF, and that
$\widehat{F}$ is therefore an estimate of the true EEF. That is to
say, $\widehat{F}$ is an estimate of the probability that a certain
measured radial distance $R$ be less than $r$:
\begin{equation}
  F(r) = P(R \le r)\,,
\end{equation}
Once we recognize the nature of the EEF, it is immediately possible to
calculate a confidence band for it by invoking the
Dvoretzky-Kieer-Wolfowitz inequality~\cite{massart90:_tight_const_dvoret_kiefer_wolfow_inequal}, that states that for
any $\epsilon > 0$
\begin{equation}
\label{eq:1}
  P\left (\sup_r|F(r) - \widehat{F}(r)| > \epsilon \right ) \le 2e^{-2n\epsilon^2}\,.
\end{equation}
If we now take~\cite{wasserman06:_all_nonpar_statis}
\begin{equation}
  \epsilon^2_n = \log(2/\alpha)/(2n)\,
\end{equation}
where $\alpha \in (0,1)$, and define
\begin{equation}
  L(r) = \max\{\widehat{F}_n(r) - \epsilon_n, 0\}
\end{equation}
and
\begin{equation}
  U(r) = \min\{\widehat{F}_n(r) + \epsilon_n, 1\}\,,
\end{equation}
by substituting in Equation~(\ref{eq:1}) we obtain
\begin{equation}
\label{eq:6}
  P\left( L(r) \le F(r) \le U(r)\right) \ge 1 - \alpha\,, \forall r\,.
\end{equation}
This equation defines for all $r$ the $1-\alpha$ confidence band for
the EEF. Examples of EEF confidence bands calculated in this manner
are shown in Figure~\ref{fig:cdf-bivariate-normal} (see
\S~\ref{sec:numer-exper} for the details).

\section{The standard error of the half energy width}
\label{sec:standard-error-half}
By definition the HEW is twice the 2-quantile (the median)
of the CDF, formally
\begin{equation}
  H = 2 F^{-1}(1/2)\,.
\end{equation}

While the confidence band calculated in the previous section can be
used to gain an idea of the uncertainty associated with the estimate
of the HEW, Equation~(\ref{eq:6}) cannot be strictly inverted: that
is, knowledge of the confidence band of the EEF cannot be used to
infer the confidence interval for the HEW. Instead we need to know the
variance and the distribution of the HEW. The bootstrap
method~\cite{efron94:_introd_boots,efron79:_boots_method} can be used
to arrive at the result.

The bootstrap is a statistical method that derives information about
the variance and the distribution of any statistics using only the
data available. Textbooks discuss the method in contexts removed from
optics: in the following we follow~\cite{efron94:_introd_boots}
(\S~5), but recast the description in terms commonly use in optics.

Going back to our observed values, $\{ R_i \}_{i=1 \ldots n}$, we
proceed as follows.
\begin{enumerate}
\item Draw with replacement from the data a new series of radial
  distances $\{R_i^\prime\}_{i=1 \ldots n}$.
\item Calculate a new HEW value $H^\prime$ from the new series.
\item Repeat the previous two steps $B$ times to obtain the bootstrap
  series of HEW estimates $\{H^\prime_k\}_{k=1 \ldots B}$.
\item The bootstrap variance estimate of the HEW is obtained by
  calculating the variance of the bootstrap series.
\end{enumerate}
The standard error we are after is the sample standard error of the
bootstrap series:
\begin{equation}
  \widehat{se} = \sqrt{\sum_{k=1}^{B} \frac{(H^\prime_k - \overline{H^\prime})^2}{B-1}}\,,
\end{equation}
where $\overline{H^\prime} = \sum_k(H^\prime_k/B)$\,. The issue is now
how large $B$ should be in order to provide a sufficiently accurate
estimate of the variance. In the literature values of $B$ of the
order of 200 are used. However a larger value is required to arrive at a good
estimate of bootstrap confidence intervals, as we discuss in the
following section.

\section{Confidence interval for the half energy width}
\label{sec:conf-interv-half}Once the standard error has been determined, we can make use of other
non-parametric techniques to estimate the confidence interval of the
HEW. In the literature a number of approaches are available to this
end. Here we discuss two of them: the percentile method, and the
so-called $\mathrm{BC}_a$ method. Both methods make use of percentiles
of the cumulative distribution of the HEW bootstrap series, but they
differ in the manner in which those percentiles are calculated. We
follow~\cite{efron87:_better_boots_confid_inter}.

\subsection{The percentile confidence interval}
\label{sec:perc-conf-interv}
The percentile method makes use of the bootstrap series of HEW
estimates to calculate the boundaries of the confidence interval. Let
$\widehat{G}$ be the empirical CDF of the bootstrap series
$\{H^\prime_k\}_{k=1 \ldots B}$ obtained above. The $1-\alpha$
percentile confidence interval for the HEW is delimited by the two
percentiles of $\widehat{G}$, $\widehat{G}^{-1}(\alpha/2)$ and
$\widehat{G}^{-1}(1 - \alpha/2)$. It is important to note that the
confidence interval obtained will not necessarily be centered on the
HEW estimate.

This method is straightforward to implement, requires $B$ to be of
the order of 2000, and makes the assumption that the bootstrap
distribution is an unbiased realization of the true HEW
distribution.

With a bit more work one can generate a better confidence interval,
better meaning that its coverage of the real confidence interval is
more accurate. In the following section we describe how this is done.

\subsection{The $\mathrm{BC}_a$ confidence interval}
\label{sec:mathrmbc_a-conf-inte}
With the $\mathrm{BC}_a$ (bias-corrected and accelerated) method the
extremes of the confidence interval are again based on two quantiles
of the bootstrap CDF, but these are calculated in a manner that
ensures that the resulting confidence interval has a higher
probability of overlapping with the true one.

The two quantiles used are:
\begin{eqnarray}
  \label{eqn:bca}
  \alpha_1 & = & \Phi\left(\hat{z}_0 + \frac{\hat{z}_0 + z^{(\alpha)}}{1 -
                 \hat{a}(\hat{z}_0 + z^{(\alpha)})}\right) \\
  \alpha_2 & = & \Phi\left(\hat{z}_0 + \frac{\hat{z}_0 + z^{(1 - \alpha)}}{1 -
                 \hat{a}(\hat{z}_0 + z^{(1 - \alpha)})}\right)\,,
\end{eqnarray}
where $\Phi(.)$ is the Normal CDF, and $\hat{a}$ and $\hat{z}_0$ are
two constants that must be computed from the data. How is explained in
detail in \S~\ref{sec:calc-mathrmbc_a-para}.

\section{Numerical experiments}
\label{sec:numer-exper}
We illustrate the issues discussed in this article with the results of
numerical simulations.

We consider the case of optics that have a point spread function that
can be represented as a circular bivariate Normal:
\begin{equation}
\label{eq:5}
  p(x,y) = \frac{1}{2\pi}\mathrm{e}^{-\frac{1}{2}(x^2 + y^2)}\,.
\end{equation}
The HEW corresponding to this function can be calculated analytically
and is
\begin{equation}
\label{eq:4}
  \mathrm{HEW} = 2\sqrt{2\ln 2} = 2r_{50}\,,
\end{equation}
where $r_{50} \approx 1.177$ is the median of the radial
distribution obtained from Equation~(\ref{eq:5}).

\begin{figure}[htbp]
  \centerline{\includegraphics[width=1.0\columnwidth]{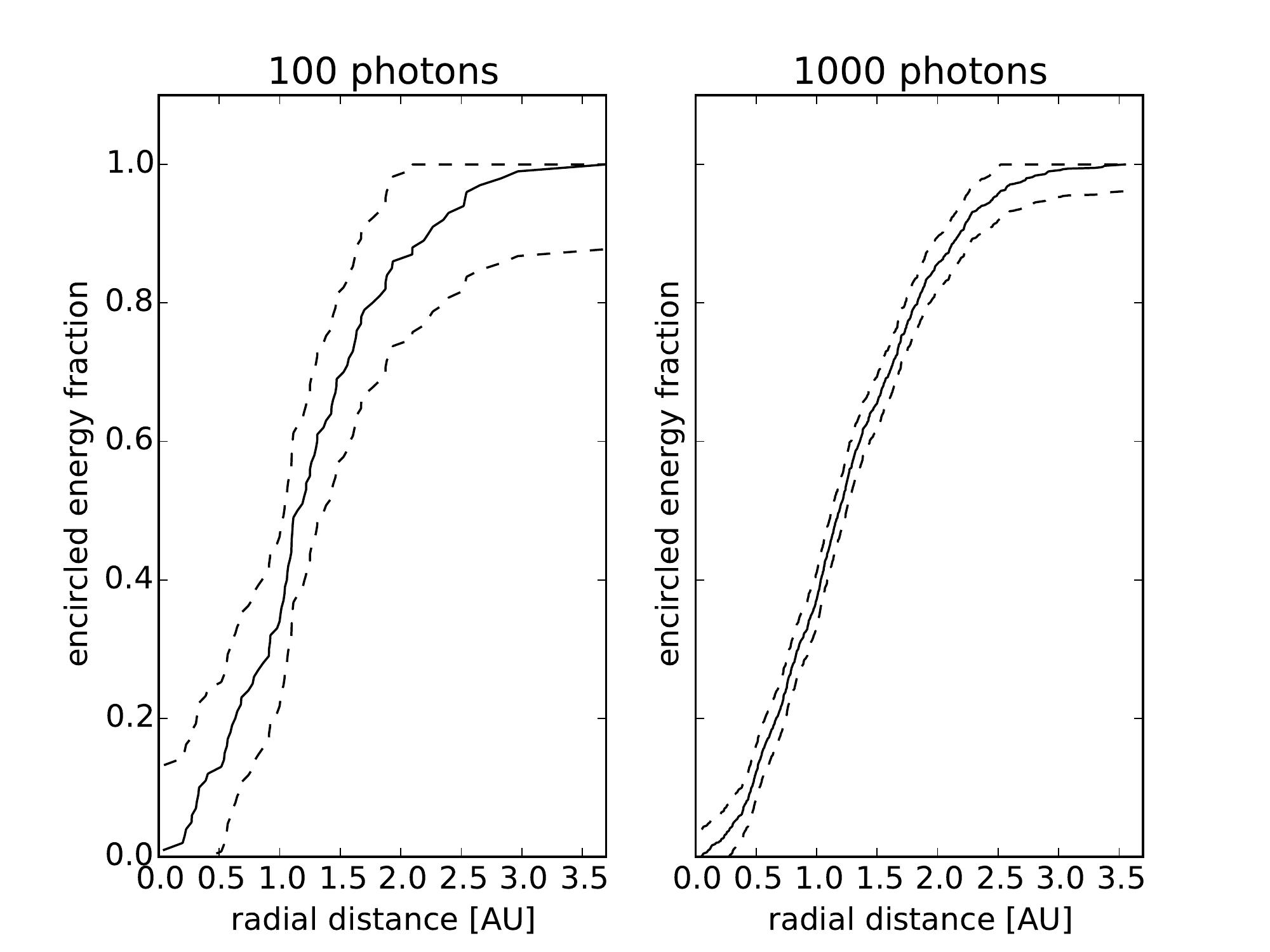}}
  \caption{Two realizations of the 90\% confidence band of the
    encircled energy fraction for an imaging system with a point
    spread function equal to a bivariate circular Normal. As can be
    expected, the confidence band narrows around the measured data
    with increasing number of photons.}
  \label{fig:cdf-bivariate-normal}
\end{figure}

In Figure~\ref{fig:cdf-bivariate-normal} we show the 90\% confidence
bands calculated with Equation~(\ref{eq:1}) for two realizations of
Equation~(\ref{eq:5}) containing 100 and 1000 photons respectively. As
can be expected the confidence band becomes narrower as the number of
photons increases. Note how relatively few photons are sufficient to
tightly bound the EEF.

\begin{figure}[htbp]
  \centering
  \includegraphics[width=1.0\columnwidth]{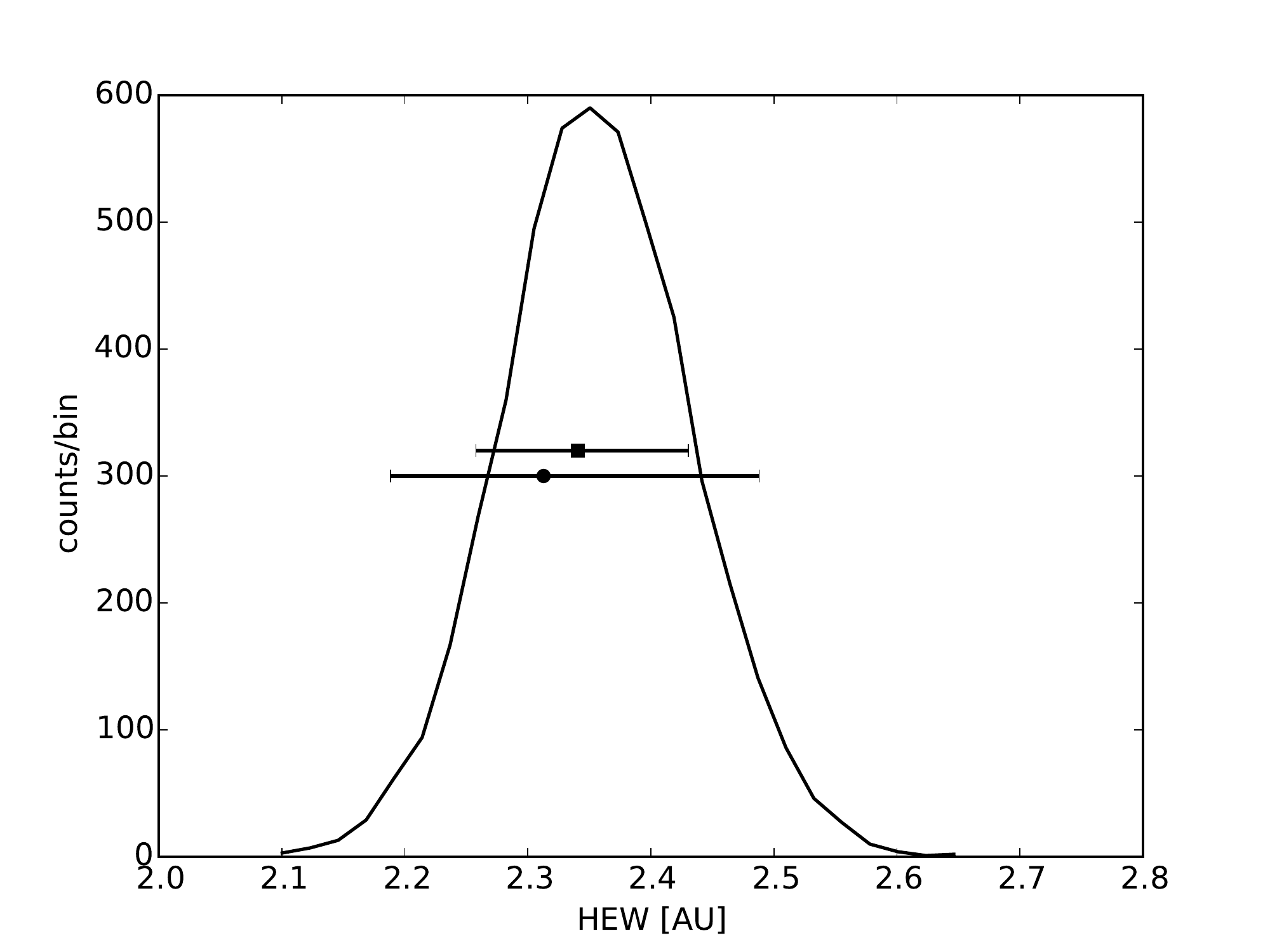}
  \caption{Simulation of a 500-photon measurement of a bivariate
    circular Gaussian point spread function. The curve shows the true
    HEW distribution, obtained by simulating 5000 measurements. Also
    shown the result of one measurement, with the 90\% confidence
    interval calculated with the $\mathrm{BC}_a$ method and
    $B=2000$. The exact HEW is about 2.35 (Equation~(\ref{eq:4}).}
  \label{fig:hew-distribution}
\end{figure}

In a second numerical experiment, we have randomly drawn 500 photons
according Equation~(\ref{eq:5}). For each of the 5000 realization we
have calculated a new HEW, and the distribution of HEW values is shown
in Figure~\ref{fig:hew-distribution}: this is meant to represent the
underlying true distribution of the infinite population of HEW
measurements. Another 500 photons were then drawn at random to
represent the measured data, and this latter dataset is used to
calculated the measured HEW and the corresponding $\mathrm{BC}_a$ 90\%
confidence interval, also shown in the figure. This example
illustrates how the distribution is indeed centered on the true value
of the HEW (Equation~(\ref{eq:4})), but that the bootstrap confidence
interval, while obviously not perfectly centered on the underlying
distribution, does indeed give a good indication of the statistical
uncertainty associated with the result.

As a final check on the validity of the approach, we remark that for a
bivariate circular Normal the radial values are distributed according
to the Rayleigh distribution, for which each quantile and its
confidence intervals can be
calculated~\cite{siddiqui1964statistical}. The HEW calculated in this
manner is also shown Figure~\ref{fig:hew-distribution}, together with
its 90\% confidence interval. The two intervals are in good agreement,
even if the confidence interval calculated analytically is smaller;
this can be easily understood because analytical methods make often
use of the knowledge of the distribution to arrive at an unbiased
estimator of minimum mean squared error. However, when the
distribution is unknown this example show that the bootstrap method
arrives at a reasonable estimate of the true confidence interval.


\section{Discussion}
\label{sec:discussion}

We have introduced methods to calculate the confidence band of the EEF
and the confidence interval of the HEW. The bootstrap method for the
computation of the confidence interval can be applied to any
measurement, but for the results to be representative of the true HEW
confidence level, the data must be taken in such a manner that they
randomly sample the point spread function of the optics. This is
certainly the case for full flood illumination measurements: then any
part of the aperture has the same probability of being probed by X-ray
photons, and the methods discussed here can be applied
straightforwardly. One could implement the method at a facility so
that the statistical accuracy reached by a measurement can be assessed
on-line, and the measurement stopped when a predefined accuracy is
arrived at. It is our opinion that in many cases this would lead to a
reduction of the required measurement time without any loss of
measurement accuracy.

We have assumed an ideal detector, or at least one with infinite
spatial resolution. Real detectors will in general be pixellated, and
may have to be read out at regularly intervals, so that the signal in
each pixel is proportional to the number of photons detected. In this
case the method can be applied by replacing Equation~(\ref{eq:7}) with
\begin{equation}
  \label{eq:9}
    \widehat{F}_n(r) = \frac{1}{\sum_1^np_i} \sum_1^n p_iI(R_i\le r)\,,
\end{equation}
where $p_i$ is the number of photons detected by the $i^\mathrm{th}$
pixel, and $R_i$ is now the radial distance of the pixel from the
optical axis.


Not all measurement methods sample the aperture of the optic
uniformly, and therefore the application of the bootstrap method
requires that changes be made to the way data are collected. Consider
for instance the characterization if Silicon Pore Optics (SPO) at the
X-ray Pencil Beam Facility of the BESSY-II
synchrotron~\cite{2013SPIE.8861E..1KV}. The SPO have an aperture that
is close to rectangular, and they are sampled with an X-ray pencil
beam following a rectangular grid. This sampling strategy does not
cover the aperture uniformly, and therefore it would have to be
changed so that the pencil beam probes the optics aperture at random
positions. Also in this case the ability to asses the statistical
error on the measured HEW is likely to deliver a significant advantage
in terms of measurement time.

In the case of optics built by assembling hundreds of individual X-ray
optics, like the Athena
mission~\cite{bavdaz10:_x_ray_pore_optic_techn}, the methods presented
here can be used to screen each individual module to a certain
accuracy before deciding whether they should be characterized further
and and how they should be eventually integrated in the larger
telescope assembly.

More in general, making measurement results available with a
statistically meaningful estimate of their significance
can only be seen as useful pursuit, and it is hoped that the
techniques described in this article may encourage the use of
statistical errors when reporting HEW results.

\section{Conclusions}
\label{sec:conclusions}

The characterization of the optical properties of X-ray optics makes
use almost exclusively of the EEF and one of its quantiles
(HEW). These are always reported without any mention of a confidence
band or confidence interval. We have shown that straightforward
non-parametric statistical methods provide ways to place a confidence
band around the EEF, and a confidence interval around the HEW.

\appendix
\section{Calculation of the $\mathrm{BC}_a$ parameters}
\label{sec:calc-mathrmbc_a-para}
In this section we show how the parameters required in the calculation
of the $\mathrm{BC}_a$ are arrived at. For the details refer
to~\cite{efron94:_introd_boots}. We consider here the original sample
$\vec{R} = \{R_i\}_{1\ldots n}$ and a statistics of interest
$\hat{\theta} = s(\vec{R})$ (in our case this would be the median or
the HEW).

The parameter $\hat{z}_0$ in Equation~\ref{eqn:bca} is called the bias
correction, and is obtained from the proportion of bootstrap estimates
smaller than the original estimate:
\begin{equation}
  \hat{z}_0 = \Phi^{-1}\left(\frac{\#\{\hat{\theta}^\prime \le \hat{\theta}\}}{B}\right)\,,
  \end{equation}
where $\Phi(.)$ is the Normal CDF.

The acceleration parameter $\hat{a}$ can be calculated in terms of the
jackknife values of the statistics $\hat{\theta}$\,. Let
$\vec{R}_{(i)}$ be the original sample with the i-th point removed,
$\hat{\theta}_{(i)} = s(\vec{R}_{(i)})$, and $\hat{\theta}_{(.)} =
\sum_1^n \hat{\theta}_{(i)} /n$\,. The acceleration parameter is
\begin{equation}
  \label{eq:2}
  \hat{a} = \frac{\sum_1^n[\hat{\theta}_{(.)} - \hat{\theta}_{(i)}]^3}{6\{\sum_1^n[\hat{\theta}_{(.)} - \hat{\theta}_{(i)}]^2\}^{2/3}}\,.
\end{equation}

\bibliographystyle{unsrt}

\bibliography{statistics-eef}

\end{document}